\documentclass[twocolumn,showpacs,preprintnumbers,amsmath,amssymb,superscriptaddress,pra]{revtex4}
\usepackage{dcolumn}
\usepackage{bm}
\usepackage{graphicx}

\newcommand{\ket}[1]{\displaystyle{|#1\rangle}}
\newcommand{\bra}[1]{\displaystyle{\langle #1|}}
\newcommand{\Ert}{\mathbf{E}(\mathbf{r},t)}
\newcommand{\Hrt}{\mathbf{H}(\mathbf{r},t)}

\newcommand{\sumkj}{\sum_{\mathbf{k}j}}
\newcommand{\sumkjp}{\sum_{\mathbf{k'}j'}}

\newcommand{\akj}{a_{\mathbf{k}j}}
\newcommand{\ackj}{a^{\dag}_{\mathbf{k}j}}
\newcommand{\akjp}{a_{\mathbf{k}'j'}}

\newcommand{\omegak}{\omega_{\mathbf{k}}}
\newcommand{\omegakp}{\omega_{\mathbf{k'}}}
\newcommand{\omegaz}{\omega_{0}}
\newcommand{\fkjrA}{\mathbf{f}_{\mathbf{k}j}(\mathbf{r}_A)}
\newcommand{\fkjprA}{\mathbf{f}_{\mathbf{k'}j'}(\mathbf{r}_A)}

\newcommand{\fkj}{\mathbf{f}_{\mathbf{k}j}}

\newcommand{\Sp}{S_{+}}
\newcommand{\Sm}{S_{-}}
\newcommand{\Sz}{S_{z}}

\newcommand{\rv}{\mathbf{r}}
\newcommand{\kv}{\mathbf{k}}
\newcommand{\ekj}{\hat{e}_{\mathbf{k}j}}
\newcommand{\ekjx}{(\hat{e}_{\mathbf{k}j})_{x}}
\newcommand{\ekjy}{(\hat{e}_{\mathbf{k}j})_{y}}
\newcommand{\ekjz}{(\hat{e}_{\mathbf{k}j})_{z}}

\def\bbm[#1]{\mbox{\boldmath $#1$}}
\DeclareMathOperator{\Ima}{Im}
\DeclareMathOperator{\Rea}{Re}
\DeclareMathOperator{\Res}{Res}

\begin{document}

\title{Time-dependent Maxwell field operators and field energy density for an atom near a conducting wall}

\author{Ruggero Vasile}
\affiliation{Department of Physics and Astronomy, University of
Turku, 20014 Turun Yliopisto, Finland}
\author{Riccardo Messina}
\affiliation{Dipartimento di Scienze Fisiche e Astronomiche
dell'Universita' degli Studi di Palermo and CNSIM, Via Archirafi 36,
I-90123 Palermo, Italy}
\affiliation{Laboratoire Kastler Brossel, case 74, CNRS, ENS, UPMC, Campus Jussieu, F-75252 Paris Cedex 05, France}
\author{Roberto Passante}
\affiliation{Dipartimento di Scienze Fisiche e Astronomiche
dell'Universita' degli Studi di Palermo and CNSIM, Via Archirafi 36,
I-90123 Palermo, Italy}

\date{\today}

\begin{abstract}
We consider the time evolution of the electric and magnetic field
operators for a two-level atom, interacting with the electromagnetic
field, placed near an infinite perfectly conducting wall. We solve
iteratively the Heisenberg equations for the field operators and
obtain the electric and magnetic energy density operators around the atom
(valid for any initial state). Then we explicitly evaluate them for
an initial state with the atom in its bare ground state and the
field in the vacuum state. We show that the results can be
physically interpreted as the superposition of the fields
propagating directly from the atom and the fields reflected on the
wall. Relativistic causality in the field propagation is discussed.
Finally we apply these results to the calculation of  the dynamical
Casimir-Polder interaction energy in the far zone between two atoms
when a boundary condition such as a conducting wall is present.
Magnetic contributions to the interatomic Casimir-Polder interaction in
the presence of the wall are also considered. We show that, in the limit of large
times, the known results of the stationary case are recovered.
\end{abstract}


\maketitle

\section{Introduction}

The existence of zero-point field fluctuations in quantum
electrodynamics has striking consequences such as Lamb shift,
spontaneous decay, Casimir and Casimir-Polder forces, which give
evidence of the quantum nature of the electromagnetic field
\cite{Milonni,CPP95}. Moreover, Casimir forces are a manifestation
of the quantum properties of the radiation field and of vacuum
fluctuations at the macroscopic level. Vacuum fluctuations are
modified by the presence of boundary conditions as well as by atoms
or molecules. Thus the presence of macroscopic objects can change
the radiative properties of the atoms. In particular, it is known
that long-range intermolecular forces change when the atoms are in
the proximity of a boundary condition such as a conducting plate or
a cavity \cite{PT82,SPR06}. This is also important from an
experimental point of view, because a precise comparison between
measurements of Casimir and Casimir-Polder forces requires
calculating these forces in situations as close as possible to
realistic laboratory situations. Also, it has been recently
suggested that appropriate boundary conditions may enhance
Casimir-Polder forces \cite{Sherkunov05,MD05,Sherkunov07,Tomas07},
and this can be very important for their direct experimental
measurement (usually they are very tiny forces).

Calculation of time-dependent electric and magnetic field operators
in the Heisenberg picture generated by an atom, even when specific boundary
conditions are present, are a powerful and general tool to calculate
various radiative processes for different initial atomic states.
This kind of calculation have been done some years ago by Power and
Thirunamachandran for atoms in the free space \cite{PT83}; they
have also successfully applied the results for the evaluation of
intermolecular Casimir-Polder forces, both for ground-state and
excited-state atoms \cite{PT83a,PT93,PT92}.

In this paper, following the same method of Power and
Thirunamachandran, we calculate the time-dependent electric and
magnetic Maxwell field operators in the Heisenberg representation generated by an
atom placed near an infinite and perfectly reflecting plate. We will
also apply our results to the evaluation of the dynamical
(time-dependent) Casimir-Polder interaction between two neutral
atoms near a conducting wall. We shall consider both electric and
magnetic interactions. Dynamical Casimir-Polder interactions have
been recently considered for atoms in the free space also for
partially dressed states \cite{PP03} or excited states \cite{RPP04},
as well as in the case of one single atom and an infinite conducting
wall \cite{SHP03,VP08,HRS04}. The method used is very suitable for generalization to other
initial states of the system.

This paper is organized as follows. In Section \ref{Sec:2} we
introduce the multipolar coupling model for a two-level atom
interacting with the radiation field in the electric dipole
approximation and in the presence of a conducting wall. Then we
solve the Heisenberg equations for the photon creation and
annihilation operators up to the second order in the electric
charge, using an iterative technique. In Sections \ref{Sec:3} we
evaluate first- and second-order corrections to the electric and
magnetic field operators, discussing their physical properties.
Finally, in Sec. \ref{Sec:4} we use the results obtained to evaluate
the electric and magnetic energy density operators and their
expectation values on the bare atom-field ground state. In Section
\ref{Sec:5} we use the expressions obtained for discussing dynamical electric
and magnetic Casimir-Polder interaction energies between two atoms
or polarizable bodies when a boundary conditions such as the
conducting plate is present. We remark that the expressions obtained
for the perturbative expansion of the fields are general and do not
depend on the specific initial state of the atom-field systems.
Hence they can be also used for evaluating Casimir-Polder energies
for different initial states such as partially dressed or excited
states.

\section{The Hamiltonian model}
\label{Sec:2}

We consider a two-level atom in front of an infinite and perfectly
conducting wall, interacting with the quantum electromagnetic
radiation field. We choose a reference frame such that the atom is
placed in $\mathbf{r}_A\equiv(0,0,d)$ whereas the mirror coincides with the
plane $z=0$. We work in the multipolar coupling scheme and within
the electric dipole approximation \cite{PowerZineau,PowerTiruna1}.
Thus the Hamiltonian describing our system reads
\begin{widetext}
\begin{equation}\begin{split}\label{IntHam}
H&=H_{0}+H_{I}\\
H_{0}&=\hbar\omegaz\Sz+\sumkj\hbar\omegak\ackj\akj\\
H_{I}&=-i\sqrt{\frac{2\pi\hbar
c}{V}}\sumkj\sqrt{k}(\bbm[\mu]\cdot\fkjrA)\bigl(\akj-\ackj\bigr)\left( \Sp + \Sm \right).\end{split}\end{equation}
\end{widetext}
In this expression the field is described by the set of bosonic
annihilation and creation operators $\akj$ and $\ackj$, associated
with a photon of frequency $\omegak=ck$, while the matrix element of the electric dipole
moment operator $\bbm[\mu]$ and the pseudospin operators $\Sp$, $\Sm$ and $\Sz$
are associated to the atom, which has a transition frequency
$\omega_0$ \cite{CPP95}. Also, $\fkj(\rv)$ are the field mode functions in the
presence of the wall, which in Eq. \eqref{IntHam} are evaluated at
the atomic position $\mathbf{r_A}$. Their expressions can be obtained
from the mode functions of a perfectly conducting cubical cavity of volume $V=L^3$ with walls ($-L/2<x,y<L/2$,
$0<z<L$) \cite{Milonni,PT82}
\begin{widetext}
\begin {equation}\label{Funzmodoparete}\begin{split}
(\fkj(\rv))_{x}
&=\sqrt{8}\ekjx\cos\Bigl[k_{x}\Bigl(x+\frac{L}2\Bigr)\Bigr]
\sin\Bigl[k_{y}\Bigl(y+\frac L2\Bigr)\Bigr]
\sin\left(k_z z\right)\\
(\fkj(\rv))_{y}
&=\sqrt{8}\ekjy\sin\Bigl[k_{x}\Bigl(x+\frac{L}2\Bigr)\Bigr]
\cos\Bigl[k_{y}\Bigl(y+\frac L2\Bigr)\Bigr]
\sin\left(k_z z\right)\\
(\fkj(\rv))_{z}&=\sqrt{8}\ekjz\sin\Bigl[k_{x}\Bigl(x+\frac{L}2\Bigr)\Bigr]
\sin\Bigl[k_{y}\Bigl(y+\frac L2\Bigr)\Bigr]
\cos\left(k_ zz\right)
\end{split}\end{equation}
\end{widetext}
where $k_x=l\pi/L$, $k_y=m\pi/L$, $k_z=n\pi/L$
($l,m,n=0,1,2,\dots$) and $\ekj$ are polarization unit vectors. In order to
switch from the cavity to the wall at $z=0$ one has to take, at the
end of the calculations, the limit $L\rightarrow\infty$.

As discussed in the Introduction, many properties of the quantum electromagnetic field and of the atom, in the presence
of the wall, can be studied in terms of the expressions of the time-dependent electric and magnetic field operators for the interacting system. In the
Heisenberg picture they can be written as
\begin{equation}\label{Campi}\begin{split}
&\Ert=i\sqrt{\frac{2\pi\hbar
c}{V}}\sumkj\sqrt{k}\,\fkj(\rv)\akj(t)+\text{h.c.}\\
&\Hrt=\sqrt{\frac{2\pi\hbar
c}{V}}\sumkj\sqrt{\frac{1}{k}}[\nabla\times\fkj(\rv)]\akj(t)+\text{h.c.}\end{split}\end{equation}
where the time dependence is fully included in the photon creation and annihilation operators. Their expressions are formally given by the solutions of the relative Heisenberg equations associated to our Hamiltonian operator \eqref{IntHam}. However, it is not possible to solve exactly these equations for our model. Thus we use an iterative technique and the solution for the annihilation operator can be written as a power series in the coupling constant
\begin{equation}\label{powerseries}
\akj(t)=\akj^{(0)}(t)+\akj^{(1)}(t)+\akj^{(2)}(t)+\dots
\end{equation}
where the contribution $\akj^{(i)}(t)$ is proportional to the $i$-th
power of the electric charge.
By taking the Hermitian conjugate we get the analogous expansion for the creation operator.
For our purposes we need their expressions up to the the second order only. The result, which is a straightforward generalization to the case with the boundary condition of that obtained for an atom in the free space \cite{PT83,RPP04}, is
\begin{widetext}
\begin{equation}\label{FotOpe}\begin{split}
\akj^{(0)}(t)&=e^{-i\omegak t}\akj\\
\akj^{(1)}(t)&=e^{-i\omegak t}\sqrt{\frac{2\pi ck}{\hbar
V}}\bigl(\bbm[\mu]\cdot\fkjrA\bigl) \bigl[\Sp F(\omegak+\omegaz,t)+\Sm
F(\omegak-\omegaz,t)\bigl]\\
\akj^{(2)}(t)&=e^{-i\omegak t}\frac{4\pi ic}{\hbar V}S_z (\bbm[\mu]\cdot\fkjrA)\sumkjp\sqrt{kk'}(\bbm[\mu]\cdot\fkjprA)\\
&\quad\times\Bigl[\akjp\Bigl(\frac{F(\omega_0+\omegak,t)-F(\omegak-\omegakp,t)}
{\omega_0+\omegakp}+\frac{F(\omegak-\omega_0,t)-F(\omegak-\omegakp,t)}{\omega_0-\omegakp}\Bigr)\\
&\quad-\akjp^\dag\Bigl(\frac{F(\omega_0+\omegak,t)-F(\omegak+\omegakp,t)}
{\omega_0-\omegakp}+\frac{F(\omegak-\omega_0,t)-F(\omegak+\omegakp,t)}{\omega_0+\omegakp}\Bigr)\Bigr]\\
\end{split}\end{equation}
\end{widetext}
where we have introduced the auxiliary function
\begin{equation}\label{FunzioniF}
F(\omega,t)=\int_{0}^{t}e^{i\omega t'}dt'=\frac{e^{i\omega t}-1}{i\omega}.
\end{equation}

All the operators appearing in the RHS of Eq. \eqref{FotOpe} are evaluated at $t=0$: from now onwards, when atomic and photonic operators are written without an explicit time dependence, it is meant they are at  $t=0$. The zeroth-order term does not depend on the presence of the source, being only the free-field contribution. As in the case of an atom in the free space discussed in  \cite{PT83}, the first order term contains only atomic operators, and the second order term contains both photonic and
atomic operators. Substituting the expansions \eqref{powerseries} and \eqref{FotOpe} into \eqref{Campi}, we get the iterative expansion of the electric and magnetic Maxwell field operators
\begin{equation}\begin{split}\label{Powerseriescampi}
&\Ert=\mathbf{E}^{(0)}(\rv,t)+\mathbf{E}^{(1)}(\rv,t)+\mathbf{E}^{(2)}(\rv,t)+\dots\\
&\Hrt=\mathbf{H}^{(0)}(\rv,t)+\mathbf{H}^{(1)}(\rv,t)+\mathbf{H}^{(2)}(\rv,t)+\dots\end{split}\end{equation}

In the next Section we explicitly evaluate the first and second order corrections of the fields.

\section{Perturbative expansion of the field operators}
\label{Sec:3}

Using the first of \eqref{FotOpe}, the expression of the zeroth-order term is
\begin{equation}\begin{split}\label{Campizero}
&\mathbf{E}^{(0)}(\rv,t)=i\sqrt{\frac{2\pi\hbar
c}{V}}\sumkj\sqrt{k}\,\fkj(\rv)e^{-i\omegak t}\akj +\text{h.c.}\\
&\mathbf{H}^{(0)}(\rv,t)=\sqrt{\frac{2\pi\hbar
c}{V}}\sumkj\sqrt{\frac{1}{k}}[\nabla\times\fkj(\rv)] e^{-i\omegak t}\akj  +\text{h.c.}
\end{split}\end{equation}

In order to obtain the first-order correction, we substitute the second of \eqref{FotOpe} into \eqref{Campi}, obtaining
\begin{equation}\begin{split}
&\mathbf{E}^{(1)}(\rv,t)=\frac{2\pi ic}{V}\sumkj k\,\fkj(\rv)\bigl(\bbm[\mu]\cdot\fkjrA\bigl)\\
&\times\biggl(\Sp F(\omegaz+\omegak,t)+\Sm
F(\omegak-\omegaz,t)\biggl)e^{-i\omegak t}+\text{h.c.}\\
&\mathbf{H}^{(1)}(\rv,t)=\frac{2\pi
c}{V}\sumkj[\nabla\times\fkj(\rv)]\bigl(\bbm[\mu]\cdot\fkjrA\bigl)\\
&\times\biggl(\Sp F(\omegaz+\omegak,t)+\Sm
F(\omegak-\omegaz,t)\biggl)e^{-i\omegak t}+\text{h.c.}
\end{split}\end{equation}

We can now take the continuum limit, by allowing $L\to \infty$. The continuum limit can be performed with the following prescription
$\sumkj\rightarrow\frac{V}{(2\pi)^3}\sum_j\int d^3\kv$. The mode functions $\fkj(\rv)$ are of course affected by this transformation since they contain $L$. The result for the $p=x,y,z$ component of the fields is
\begin{widetext}
\begin{equation}\begin{split}
E_p^{(1)}(\rv,t) &=\frac{ic}{4\pi^2}\sum_{q}\mu_{q}\int dk\,k^3\Bigl(\Sp F(\omegaz+\omegak,t)+\Sm F(\omegak-\omegaz,t)\Bigr)e^{-i\omegak
t}\sum_{j}\int d\Omega(\fkj(\rv))_{p}(\fkjrA)_{q}\\
H_p^{(1)}(\rv,t) &=\frac{c}{4\pi^2}\sum_{q}\mu_{q}\int dk\,k^2\Bigl(\Sp
F(\omegaz+\omegak,t)+\Sm F(\omegak-\omegaz,t)\Bigr)e^{-i\omegak t}\sum_{j}\int
d\Omega(\nabla\times\fkj(\rv))_{p}(\fkjrA)_{q}
\end{split}
\end{equation}
\end{widetext}
where the integrations over $k$ run from $-\infty$ to $+\infty$,
$\mu_q$ is the $q=x,y,z$ component of $\bbm[\mu]$. The polarization
sum and the integral over angular variables can be explicitly
calculated by lengthy straightforward calculations (see the Appendix
for more details). Finally, we get

\begin{widetext}
\begin{equation}\begin{split}
E_p^{(1)}(\rv,t)
&=\frac{ic}{\pi}\sum_q\sum_{\nu=0,1}(-1)^{\nu(1-\delta_{qz})}
\mu_qF_{pq}\frac{1}{r_{\nu}}\int dk\,\Bigl(\Sp
F(\omegaz+\omegak,t)+\Sm
F(\omegak-\omegaz,t)\Bigr)e^{-i\omegak t}\sin{(kr_{\nu})}\\
H_p^{(1)}(\rv,t)&=\frac{c}{\pi}\sum_q\sum_{\nu=0,1}
(-1)^{\nu(1-\delta_{qz})} \mu_qG_{pq}\frac{1}{r_{\nu}}\int
dk\,k\Bigl(\Sp F(\omegaz+\omegak,t)+\Sm
F(\omegak-\omegaz,t)\Bigr)e^{-i\omegak t}\sin(kr_{\nu})
\end{split}\end{equation}
\end{widetext}
where $F_{pq}=-\nabla^2\delta_{pq}+\nabla_p\nabla_{q}$,
$G_{pq}=-\epsilon_{pqs}\nabla_s$ ($\epsilon_{\alpha\beta\gamma}$
being the complete antisymmetric tensor), and
$\nabla_p\equiv\frac{\partial}{\partial x_p}$ ($p=1,2,3$) are
differential operators acting on the variables $x,y,z$. The index
$\nu=0,1$ labels two possible values of the quantity $r_{\nu}$,
namely $r_0=|\rv_0|=|\rv-\mathbf{r}_A|=\sqrt{x^2+y^2+(z-d)^2}$ and
$r_1=|\rv_1|=|\rv-\mathbf{r}_I|=\sqrt{x^2+y^2+(z+d)^2}$, where
$\mathbf{r}_I\equiv(0,0,-d)$ is the position of the image of the
atom reflected on the wall. $r_0$ is the distance of the observation
point $\rv = (x,y,z)$ from the atom in $\mathbf{r}_A\equiv(0,0,d)$,
whereas $\mathbf{r}_I\equiv(0,0,-d)$ is its distance from the atom's
image.

The final step for the evaluation of the first-order correction is
the evaluation of the $k$-integral. Using  \eqref{FunzioniF} and the
following integral representations of the Heaviside step function
$\theta(x)$
\begin{equation}\begin{split}\label{theta}
&\int_{-\infty}^\infty dk\,\frac{1-e^{\pm
i(k_{0}-k)ct}}{k-k_{0}}\sin(kr_{\nu})=\pi e^{\pm
ik_{0}r_{\nu}}\theta(ct-r_{\nu})\\
&\int_{-\infty}^\infty dk\,\frac{1-e^{ i(k\pm k_{0})ct}}{k\pm
k_{0}}\cos(kr_{\nu})=-i\pi e^{\pm ik_{0}r_{\nu}}\theta(ct-r_{\nu})
\end{split}\end{equation}
we obtain the following expressions for the first-order correction of the
electric and magnetic field operators generated by the two-level atom
in front of a perfectly conducting wall
\begin{widetext}
\begin{equation}\label{FinPri}\begin{split}
E_p^{(1)}(\rv,t)&=\sum_q\sum_{\nu=0}^{1}(-1)^{\nu(1-\delta_{qz})}
\mu_{q}F_{pq}\frac{1}{r_\nu}\bigl[\Sp e^{ik_{0}(ct-r_{\nu})}+\Sm
e^{-ik_{0}(ct-r_{\nu})}\bigl]\theta(ct-r_{\nu})\\
H_p^{(1)}(\rv,t)
&=-\sum_q\sum_{\nu=0}^{1}(-1)^{\nu(1-\delta_{qz})}
\mu_{q}G_{pq}\frac{1}{r_{\nu}}\frac{\partial}{\partial r_{\nu}}\bigl[\Sp  e^{ik_{0}(ct-r_{\nu})}+\Sm
e^{-ik_{0}(ct-r_{\nu})}\bigl]\theta(ct-r_{\nu}).
\end{split}\end{equation}
\end{widetext}
We wish to stress that expressions \eqref{FinPri} are general operatorial relations, which can be used for calculating matrix elements between any pair of atom-field states.

It is worthy analyzing from a physical point of view the expressions
obtained. The sum over $\nu$  in \eqref{FinPri} describes the
existence of two distinct contributions to these operators. The
first one depends on the parameter $r_0 = |\rv-\mathbf{r}_A|$, that
is the distance between the  observation point $\rv$ and the atom.
We shall refer to it as the \emph{direct} term. The second one is a
function of $r_1 = |\rv-\mathbf{r}_I|$, that is the distance between
the observation point and the image atom. We will call it the
\emph{reflected} term. Both terms propagate satisfying rigorously
the causality principle as shown by the presence of the $\theta$
functions. We can check that our results \eqref{FinPri} correctly
yield the known analogous results for an atom in the free space as
obtained in \cite{PT83}, if we take the limiting case  of the wall
very distant from the atom: in fact, in this limit $r_0$ is finite,
whereas $r_1$ grows with the atom-wall distance. In the same limit
we can thus keep only the $\nu=0$ term  in \eqref{FinPri}, and obtain the
same expressions found in \cite{PT83}. A final remark concerning the
action of the differential operators in \eqref{FinPri} is necessary,
because in principle they should be applied to the Heaviside
functions also. This means that the Dirac delta function and its
derivatives show up, yielding a singular impulse propagating as a
wave front on the light cone, strongly dependent on the model used
(in particular on the dipole approximation). Similar singular
behaviour were found in the field emitted during the dynamical
dressing of an atom \cite{CPP88,VP08,PP03}. In the applications we
will discuss in the following Sections we will not consider the
electromagnetic field on this (singular) wave front and so we will
not explicitly evaluate it (hence the results in the next Section
will not be valid for $t=r_0/c$ or $t=r_1/c$).

We now consider the second-order correction to the electric and
magnetic field operators, obtained by substitution of  the third of
\eqref{FotOpe} into \eqref{Campi}. Since the calculation is quite
straightforward, although lengthy, we shall only give the final results. In the expressions obtained, two
independent sums over wavevectors and polarizations are present: one
of them can be explicitly performed and finally yields a $\theta$ function
ensuring relativistic causality. Thus we get

\begin{widetext}
\begin{equation}\label{Finsec}\begin{split}
E_p^{(2)}(\rv,t)&=-2i\Sz \sqrt{\frac{2\pi c}{\hbar
V}}\sumkj\sqrt{k}\bigl(\bbm[\mu]\cdot\fkjrA\bigl)\sum_q\sum_{\nu=0,1}(-1)^{\nu(1-\delta_{qz})}
\mu_{q}
F_{pq}\frac{1}{r_\nu}\\
\times\biggl\{&\akj \biggl[\frac{1}{\omegaz+\omegak}\bigl(e^{-ik(ct-r_\nu)}-e^{ik_{0}(ct-r_\nu)}\bigl)
+\frac{1}{\omegaz-\omegak}\bigl(e^{-ik(ct-r_\nu)}-e^{-ik_{0}(ct-r_\nu)}\bigl)\biggl]+\\
-&\ackj  \biggl[\frac{1}{\omegaz+\omegak}\bigl(e^{ik(ct-r_\nu)}-e^{-ik_{0}(ct-r_\nu)}\bigl)
+\frac{1}{\omegaz-\omegak}\bigl(e^{ik(ct-r_\nu)}-e^{ik_{0}(ct-r_\nu)}\bigl)\biggl]\biggl\}\theta(ct-r_\nu)\\
H_p^{(2)}(\rv,t)&=2i\Sz \sqrt{\frac{2\pi c}{\hbar
V}}\sumkj\sqrt{k}\bigl(\bbm[\mu]\cdot\fkjrA\bigl)\sum_q\sum_{\nu=0,1}\mu_{q}(-1)^{\nu(1-\delta_{qz})}G_{pq}
\frac{1}{r_{\nu}}\\
\times\frac{\partial}{\partial r_{\nu}}\biggl\{\akj \biggl[&\frac{1}{\omegaz+\omegak}\bigl(e^{-ik(ct-r_{\nu})}
-e^{ik_{0}(ct-r_{\nu})}\bigl)+\frac{1}{\omegaz-\omegak}
\bigl(e^{-ik(ct-r_{\nu})}-e^{-ik_{0}(ct-r_{\nu})}\bigl)\biggl]\\
-\ackj  \biggl[&\frac{1}{\omegaz+\omegak}\bigl(e^{ik(ct-r_{\nu})}-e^{-ik_{0}(ct-r_{\nu})}\bigl)
+\frac{1}{\omegaz-\omegak}\bigl(e^{ik(ct-r_{\nu})}-e^{ik_{0}(ct-r_{\nu})}\bigl)\biggl]\biggl\}\theta(ct-r_{\nu}).
\end{split}\end{equation}
\end{widetext}

The second-order fields in \eqref{Finsec}, similarly to the
first-order ones \eqref{FinPri}, show the presence of a direct and a
reflected term.

\section{Electric and magnetic energy densities of the field}
\label{Sec:4}

In this Section we shall use the results of Section \ref{Sec:3} for
calculating the electric and magnetic field energy densities
generated by the atom when the plate is present; they are of
course related to the field fluctuations generated by the atom. In
the next Section the results obtained will be applied  to
evaluate the dynamical Casimir-Polder potential between two atoms
placed near the conducting plate (in the so-called far zone).

We shall first evaluate the square of the electric field operator and
its average value on the initial state we are mostly interested in.
The square of the electric field operator coincides with the electric energy
density operator, apart a factor
$1/8\pi$, and also with the electric field fluctuations, assuming a
vanishing expectation value of the electric field. Using the expansion of the electric field up to
the second order, we have

\begin{equation}\begin{split}\label{Eledens}
&E^2(\rv,t)=\mathbf{E}^{(0)}(\rv,t)\cdot\mathbf{E}^{(0)}(\rv,t)\\
&\quad+\Bigl(\mathbf{E}^{(0)}(\rv,t)\cdot\mathbf{E}^{(1)}(\rv,t)+\mathbf{E}^{(1)}(\rv,t)\cdot\mathbf{E}^{(0)}(\rv,t)\Bigr)\\
&\quad+\Bigl(\mathbf{E}^{(1)}(\rv,t)\cdot\mathbf{E}^{(1)}(\rv,t)+\mathbf{E}^{(0)}(\rv,t)\cdot\mathbf{E}^{(2)}(\rv,t)\\
&\quad+\mathbf{E}^{(2)}(\rv,t)\cdot\mathbf{E}^{(0)}(\rv,t)\Bigr)+\dots\end{split}\end{equation}

The first term is a free-field term that does not depend on the presence of the source; then we have first- and second-order terms. From now onwards we consider the specific initial atom-field state $\ket{0_{\mathbf{k}j},\downarrow_A}$, where the field is in its vacuum state and the ``source'' atom (A) is in its bare ground state, and calculate the (time-dependent) expectation value of \eqref{Eledens}.  This quantity gives the time evolution of the electric energy density during the self-dressing process of atom A, that in the next Section will be used to evaluate the dynamical Casimir-Polder interaction energy between two atoms when the conducting plate is present.

The average value of the zeroth-order contribution is independent of the presence of the atom and thus does not contribute to atom-atom interactions (it gives the time-independent but spatially dependent fluctuations of the electric field with the plate present and in the absence of the atom). Also, the first-order term has a zero mean value on our initial state. Thus, the first relevant non-vanishing contributions are given by the second-order terms. We start evaluating the average value of $\mathbf{E}^{(1)}(\rv,t)\cdot\mathbf{E}^{(1)}(\rv,t)$
\begin{widetext}
\begin{equation}\begin{split}
&\bra{0_{\mathbf{k}j},\downarrow_A}
\mathbf{E}^{(1)}(\rv,t)\cdot\mathbf{E}^{(1)}(\rv,t)
\ket{0_{\mathbf{k}j},\downarrow_A}  \\
&\quad=\sum_{pqq'}\sum_{\nu\nu'=0,1}\mu_{q}\mu_{q'}(-1)^{\nu(1-\delta_{qz})}(-1)^{\nu'(1-\delta_{q'z})}
F_{pq}\biggl(\frac{e^{ik_0r_{\nu}}}{r_{\nu}}\theta(ct-r_{\nu})\biggl)
F_{pq'}\biggl(\frac{e^{-ik_0r_{\nu'}}}{r_{\nu'}}\theta(ct-r_{\nu'})\biggl)
\end{split}\end{equation}
\end{widetext}
where $k_0=\omegaz /c$. As discussed in Section \ref{Sec:3}, we do not to apply the differential operators to the $\theta$ functions, because this would only add singular terms on the light cone, in which we are not interested at the moment. Using the result
\begin{equation}
F_{pq}\frac{e^{ikr}}{r}=k^3f_{pq}(kr)e^{ikr}
\end{equation}
where
\begin{equation}\label{Tens}
f_{pq}(kr)=\frac{\delta_{pq}-\hat{r}_p\hat{r}_q}{kr}+(\delta_{pq}-3\hat{r}_p\hat{r}_q)\Bigl(\frac{i}{k^2r^2}-\frac{1}{k^3r^3}\Bigl)\\
\end{equation}
we obtain
\begin{widetext}
\begin{equation}\label{E1E1}\begin{split}
\bra{0_{\mathbf{k}j},\downarrow_A}
\mathbf{E}^{(1)}(\rv,t)\cdot\mathbf{E}^{(1)}(\rv,t)
\ket{0_{\mathbf{k}j},\downarrow_A}
&=-\sum_{\nu\nu'=0,1}k_0^6\mu_{q}\mu_{q'}(-1)^{\nu(1-\delta_{qz})}(-1)^{\nu'(1-\delta_{q'z})}
f_{pq}(k_0r_{\nu})f_{pq'}(-k_0r_{\nu'})\\
&\quad\times e^{ik_0(r_{\nu}-r_{\nu'})}\theta(ct-r_{\nu})\theta(ct-r_{\nu'}).
\end{split}\end{equation}
We can analyze the previous expression in three different space-time regions. When $ct<r_0$, that is for times shorter than the time taken by a light signal to travel from the atom in $\rv_A$ to the observation point $\rv$, it is zero because both Heaviside functions are vanishing. For $r_0<ct<r_1$, i.e. after that a light signal can travel from the atom to the observation point but before this point has been reached by the signal reflected on the wall, only the contribution with $\nu=\nu'=0$ is different from zero and the result is the same obtained in \cite{PT92} for an atom in the free space. For $ct>r_1$, that is after that even the reflected signal has passed the observation point, all terms contribute, and we can recognize a contribution from the atom ($\nu=\nu'=0$), one from the image atom ($\nu=\nu'=1$) and also interference terms ($\nu\neq\nu'$). The result in free space is recovered by taking the limit $r_0<<r_1$ (i.e. very distant wall). Similar consideration holds for the other contributions we are now going to evaluate.

In the continuum limit, the contribution from second-order fields  in \eqref{Eledens} becomes
\begin{equation}\begin{split}
&\bra{0_{\mathbf{k}j},\downarrow_A}\mathbf{E}^{(0)}(\rv,t)\cdot\mathbf{E}^{(2)}(\rv,t)\ket{0_{\mathbf{k}j},\downarrow_A}
=-\frac{1}{\pi}\sum_{pqq'}\sum_{\nu\nu'=0,1}\mu_{q}\mu_{q'}
(-1)^{\nu(1-\delta_{qz})}(-1)^{\nu'(1-\delta_{q'z})}\theta(ct-r_{\nu})\\
&\quad\times\int_{0}^{+\infty}dk\,e^{-ikct} F_{pq}\frac{1}{r_{\nu}}\biggl[\frac{1}{k+k_0}
\bigl(e^{-ik_{0}(ct-r_\nu)}-e^{ik(ct-r_\nu)}\bigl)+\frac{1}{k_0-k}\bigl(e^{ik_{0}(ct-r_\nu)}-e^{ik(ct-r_\nu)}\bigl)\biggl]
F_{pq'}\biggl[\frac{\sin(kr_{\nu'})}{r_{\nu'}}\biggl].
\end{split}\end{equation}

Finally, after some straightforward algebra, we obtain
\begin{equation}\begin{split}\label{E2E0}
&\bra{0_{\mathbf{k}j},\downarrow_A}\mathbf{E}^{(0)}(\rv,t)\cdot\mathbf{E}^{(2)}(\rv,t)\ket{0_{\mathbf{k}j},\downarrow_A} +
\bra{0_{\mathbf{k}j},\downarrow_A}\mathbf{E}^{(2)}(\rv,t)\cdot\mathbf{E}^{(0)}(\rv,t)\ket{0_{\mathbf{k}j},\downarrow_A} \\
&\quad=-\frac{1}{2\pi i}\sum_{p,q,q'}\sum_{\nu,\nu'=0}^1\int_0^{+\infty}dk\,k^3\mu_q\mu_{q'}(-1)^{\nu(1-\delta_{qz})}(-1)^{\nu'(1-\delta_{q'z})}
\theta(ct-r_\nu)\\
&\qquad\times\Bigl[\frac{1}{k_0+k}\Bigl(k_0^3f_{pq}(k_0r_\nu)f_{pq'}(kr_{\nu'})e^{-ik_0(ct-r_\nu)}e^{-ik(ct-r_{\nu'})}+\\
&\qquad+k_0^3f_{pq}(k_0r_\nu)f_{pq'}(-kr_{\nu'})e^{-ik_0(ct-r_\nu)}e^{-ik(ct+r_{\nu'})}+k^3f_{pq}(-kr_\nu)f_{pq'}(-kr_{\nu'})
e^{-ik(r_\nu+r_{\nu'})}\Bigr)\\
&\qquad+\frac{1}{k_0-k}\Bigl(-k_0^3f_{pq}(-k_0r_\nu)f_{pq'}(kr_{\nu'})e^{ik_0(ct-r_\nu)}e^{-ik(ct-r_{\nu'})}\\
&\qquad-k_0^3f_{pq}(-k_0r_\nu)f_{pq'}(-kr_{\nu'})e^{ik_0(ct-r_\nu)}e^{-ik(ct+r_{\nu'})}+k^3f_{pq}(-kr_\nu)f_{pq'}(-kr_{\nu'})
e^{-ik(r_\nu+r_{\nu'})}\Bigr)\Bigr]+\text{c.c.}\\\end{split}\end{equation}
\end{widetext}
where the terms with a pole in $k=k_0$ should be taken as their
principal value. We now extend the integral over $k$ to the complex
plane, closing the integration path at infinity in the fourth
quadrant and on the negative imaginary axis. Assuming no other
singularity in the integrand function except the pole at $k=k_0$ and that it
vanishes at infinity, the result can be written as an integral along
the imaginary axis plus the pole contribution
\begin{equation} \label{complint}
P\int_0^\infty dk\,g(k)=-i\int_0^\infty dk\,g(-ik)-i\pi\Res\{g(k)\}_{k=k_0}
\end{equation}
where $g(k)$ indicates the integrand function.

We now give the explicit expression of
the expectation value of the square of the electric field for
$t>r_1/c$, that is after that the direct and reflected waves have
reached the observation point (this quantity vanishes for $t<r_0/c$,
of course, but it is not zero for $r_0/c<t<r_1/c$ due to the ``direct" contribution).
In this timescale, the contribution from
the pole at $k=k_0$ for $t>r_1/c$ in \eqref{complint} exactly cancels the contribution \eqref{E1E1}
when the three second-order contributions in \eqref{Eledens} are
summed up, exactly as in the case of an atom in the free space
discussed in \cite{PT92}.  We obtain
\begin{widetext}
\begin{equation}\begin{split}\label{Equadro}
\bra{0_{\mathbf{k}j},\downarrow_A}E^2(\rv,t)\ket{0_{\mathbf{k}j},\downarrow_A}&-E_{\text{zp}}^E=-\frac{\hbar c}\pi \sum_{pqq'}\sum_{\nu\nu'=0}^{1}(-1)^{\nu(1-\delta_{qz})}(-1)^{\nu'(1-\delta_{q'z})}\\
&\times\Biggl[\int_{0}^\infty dk\,k^6\alpha_{qq'}(ik)f_{pq}(ikr_{\nu})f_{pq'}(ikr_{\nu'})e^{-k(r_{\nu}+r_{\nu'})}\\
&-\frac{ik_0^2}2
\int_0^\infty dk\,k^3e^{-kct}\alpha_{qq'}(ik)\Bigl(k_0\Rea f_{pq}(k_0r_{\nu})e^{-ik_0(ct-r_{\nu})}-k\Ima
f_{pq}(k_0r_{\nu})e^{-ik_0(ct-r_{\nu})}\Bigr)\\
&\times\Bigl(f_{pq'}(-ikr_{\nu'})e^{kr_{\nu'}}+f_{pq'}(ikr_{\nu'})e^{-kr_{\nu'}}\Bigr)\Biggr]\end{split}\end{equation}
\end{widetext}
where we have explicitly indicated that the ``zero-point" electric
energy density $E_{\text{zp}}^E$ due to zero-order field operators has
been subtracted, and we have also introduced the dynamic atomic
polarizability tensor at imaginary frequencies
\begin{equation}
\alpha_{qq'}(ik)=\frac{2k_0\mu_q\mu_{q'}}{\hbar c(k^2+k_0^2)}.
\end{equation}
It is worth considering the limiting case of Eq. \eqref{Equadro} for $t\to+\infty$. In this limit, fast oscillating terms inside the
$k$-integral average to zero, finally yielding the time-independent value given by the first two lines only of Eq.  \eqref{Equadro}.
For an isotropic atom or polarizable body ($\mu_x=\mu_y=\mu_z$) this becomes
\begin{widetext}\begin{equation}\label{Enind}\begin{split}
&\bra{0_{\mathbf{k}j},\downarrow_A}E^2(\rv)\ket{0_{\mathbf{k}j},\downarrow_A} -E_{\text{zp}}^E=\frac{2\hbar c}{\pi}\sum_{\nu=0}^{1}\int_0^{+\infty}dk\,k^6\alpha(ik)\Bigl(\frac{1}{k^2r_{\nu}^2}+\frac{2}{k^3r_{\nu}^3}+
\frac{5}{k^4r_{\nu}^4}+\frac{6}{k^5r_{\nu}^5}+\frac{3}{k^6r_{\nu}^6}\Bigr)e^{-2kr_{\nu}}\\
&\quad+\frac{2\hbar c}{\pi}\sum_{pq}(-1)^{1-\delta_{qz}}\int_0^{+\infty}dk\,k^6\alpha(ik)\Bigl[\frac{\delta_{pq}-\hat{r}_{0p}\hat{r}_{0q}}{kr_0}+(\delta_{pq}-3\hat{r}_{0p}\hat{r}_{0q})
\Bigl(\frac{1}{k^2r_0^2}+\frac{1}{k^3r_0^3}\Bigr)\Bigr]\\
&\quad\times\Bigl[\frac{\delta_{pq}-\hat{r}_{1p}\hat{r}_{1q}}{kr_1}+(\delta_{pq}-3\hat{r}_{1p}\hat{r}_{1q})\Bigl(\frac{1}{k^2r_1^2}+\frac{1}{k^3r_1^3}\Bigr)\Bigr]e^{-k(r_0+r_1)}.\\
\end{split}\end{equation}\end{widetext}
where $\alpha(k)$ is the isotropic electric polarizability of the atom. There are three different terms that contribute to the electric energy density. The first comes from the $\nu=\nu'=0$ contribution and it is identical to that found in \cite{PT92} for a two-level atom in the free space. The second term is the $\nu=\nu'=1$ contributions and has the same functional expression of the previous one with $r_0$ replaced by $r_1$ (distance of the observation point from the image of the atom). Finally, there is a \emph{mixed} term depending on both distances $r_0$ and $r_1$. The physical meaning and role on
interatomic Casimir-Polder interactions of these terms will be discussed in detail in the next Section.

Similarly to the electric case, from the results of Section \ref{Sec:3} we can calculate the magnetic energy density, which we shall use in the next Section in order to calculate magnetic interatomic Casimir forces in the far zone. We shall not give the explicit expressions of the time-dependent magnetic energy density, but only its asymptotic expression for $t\to+\infty$. In this limit it settles to a time-independent value, as the electric energy density, given by
\begin{widetext}
\begin{equation}\begin{split}
&\bra{0_{\mathbf{k}j},\downarrow_A}H^2(\rv)\ket{0_{\mathbf{k}j},\downarrow_A} -E_{\text{zp}}^M=\frac{2k_0}{\pi}\sum_{p,q,q'}
\sum_{\nu,\nu'=0,1}\mu_q\mu_{q'}(-1)^{\nu(1-\delta_{qz})}(-1)^{\nu'(1-\delta_{q'z})}\\
&\qquad\times\int_0^{+\infty}dk\,\frac{k^6}{k_0^2+k^2}g_{pq}(ikr_\nu)g_{pq'}(ikr_{\nu'})e^{-k(r_\nu+r_{\nu'})}\\
&\quad=\frac{\hbar c}{\pi}\sum_{p,q,q'}\sum_{\nu,\nu'=0,1}(-1)^{\nu(1-\delta_{qz})}(-1)^{\nu'(1-\delta_{q'z})}\int_0^{+\infty}dk\,k^6\alpha_{qq'}(ik)g_{pq}(ikr_\nu)g_{pq'}(ikr_{\nu'})e^{-k(r_\nu+r_{\nu'})}\\
\end{split}\end{equation}
(similarly to the electric case, the zero-point magnetic energy density has been subtracted) where
\begin{equation}
g_{pq}(ikr)=i\sum_s\epsilon_{pqs}\hat{r}_s\Bigl(\frac{1}{kr}+\frac{1}{k^2r^2}\Bigr).
\end{equation}
For an isotropic molecule it becomes
\begin{equation}\label{Hquadro}
\bra{0_{\mathbf{k}j},\downarrow_A}H^2(\rv)\ket{0_{\mathbf{k}j},\downarrow_A} -E_{\text{zp}}^M=\frac{\hbar c}{\pi}\sum_{p,q}\sum_{\nu,\nu'=0,1}(-1)^{(\nu+\nu')(1-\delta_{qz})}\int_0^{+\infty}dk\,k^6\alpha(ik)g_{pq}(ikr_\nu)g_{pq}(ikr_{\nu'})e^{-k(r_\nu+r_{\nu'})}.
\end{equation}\end{widetext}

\section{The far-zone electric and magnetic interatomic Casimir-Polder energies with boundary conditions}
\label{Sec:5}

In this Section we apply the results of the previous Sections for the time-dependent Maxwell fields to the calculation of dynamical electric and magnetic Casimir-Polder interactions between two atoms in the presence of the conducting wall, also recovering known results in the asymptotic regime. The electric and magnetic field operators of Section \ref{Sec:3}, and the related expectation values of the energy densities obtained in Section \ref{Sec:4}, can be used to obtain static and dynamical Casimir-Polder forces between two atoms in the presence of a boundary condition such as a conducting wall. In fact, the
interatomic Casimir-Polder interaction in the so-called far zone, that is for distances between the atoms larger than their typical transition wavelength from the ground state, can be obtained from the interaction of one atom, described as a polarizable body, with the field fluctuations generated by the other atom \cite{PP87,CP69,PPT98}. This approach has been recently proved to be useful also in the case of dynamical (time-dependent) Casimir-Polder forces for atoms in excited states \cite{RPP04} or in partially dressed states \cite{PP03}. Thus we evaluate the (electric)
Casimir-Polder interaction energy in the far zone between our ``source'' atom A and a second atom B placed in $\rv_B$, when the conducting wall in present, as
\begin{equation}\label{ElIntEn}
\Delta E_E = - \frac 12 \alpha_E^B \langle E^2(\rv_B ,t) \rangle_A
\end{equation}
where $\alpha_E^B$ is the static electric polarizability of atom B, assumed isotropic for simplicity, and the average value of $E^2(\rv,t)$, evaluated at the position of atom B, is calculated on the initial state of the field and of atom A (for example the bare ground state, as in the previous Section). In a quasi-static approach, the Casimir-Polder force is then obtained as minus the derivative of the interaction energy with respect to the distance between the two atoms. Effects related to the atomic motion, as discussed in \cite{SHP03,HRS04}, are neglected here. Besides being well suited for dealing with dynamical interactions, our method also
allows a simpler derivation of static Casimir-Polder forces and a transparent physical interpretation, compared to the usual methods based on fourth-order perturbation calculations \cite{CraigThiru}.

From our results dynamical magnetic Casimir-Polder interactions can also be easily obtained, such as the interaction of a magnetically polarizable body B (with static magnetic polarizability $\alpha_B^M$) with the magnetic field fluctuations generated by atom A, as
\begin{equation}\label{MagIntEn}
\Delta E_M = - \frac 12 \alpha_M^B \langle H^2(\rv_B ,t) \rangle_A.
\end{equation}

The relations above show that the Casimir-Polder interaction energy between two atoms is proportional to the expectation value of the square of the fields generated by one of the two atoms at the position of the other atom, that is to the average electric (and magnetic) energy density or equivalently to the field fluctuations generated by the atom.

In order to evaluate equation \eqref{ElIntEn} in the far zone, we can use the expression \eqref{Enind} of the electric energy density with $\rv =\rv_B$ (in the limit $t \to \infty$), approximated by substituting the dynamical polarizability of the atom with the static one; this allows to perform the $k$ integral. We finally obtain
\begin{equation} \begin{split}\label{ElCP}
\Delta E_E&=-\frac{1}{2}\alpha_E^B\langle E^2(\rv_B)\rangle_A
=-\frac{23\hbar c}{4\pi}\alpha_E^A\alpha_E^B\Bigl(\frac{1}{r_0^7}+\frac{1}{r_1^7}\Bigl) \\
&\quad+\frac{16\hbar c\alpha_E^A\alpha_E^B}{\pi r_0^3r_1^3(r_0+r_1)^5}\Bigl[3r_0^2r_1^2+\rho^2(r_0^2+5r_0r_1+r_1^2)\Bigr]
\end{split}\end{equation}
with
\begin{equation}
\rho=r_0\sin\theta_0=r_1\sin\theta_1
\end{equation}
where $\theta_0$ ($\theta_1$) is the angle between the vector $\rv_0$ ($\rv_1$) and the perpendicular to the wall.

Expression \eqref{ElCP}, which was already obtained with different methods in \cite{PT82,SPR06}, shows that in the presence of the conducting plate, the Casimir-Polder interaction between the two atoms A and B consists of three terms: the $r_0^{-7}$ interaction between the two atoms as in the absence of the wall; the $r_1^{-7}$ interaction between one atom and the image of the other atom as reflected by the wall; a term involving both coordinates $r_0$ and $r_1$. The first two terms give an attractive component, whereas the third one gives a repulsive component. However, as discussed in \cite{SPR06}, the total interaction is attractive for any geometric configuration of the atoms with respect to the wall.

Our results in the previous Section also allow to evaluate the time-dependent Casimir-Polder interaction between the two atoms, using the time-dependent expression for the squared electric energy density \eqref{Equadro} (and not only its asymptotic value for $t\to\infty$). The interaction energy between the two atoms in the far zone as a function of time for $ct>r_1$, that is after that both the direct and reflected fields generated by atom A have reached atom B, is characterized by oscillations around its asymptotic value, with an amplitude decaying with time to zero. For $ct\gg r_1$ the energy settles to its stationary value, as obtained in a time-independent approach \cite{SPR06} or from \eqref{ElCP}. This behavior is analogous to that already found in the time-dependent Casimir-Polder force between an atom and a conducting wall \cite{VP08}.

In a similar way we can explicitly evaluate the magnetic far-zone Casimir-Polder interaction energy,  in the presence of the wall, by substituting \eqref{Hquadro} into  \eqref{MagIntEn}. In the far zone, after substitution of the dynamical polarizability of the atom with the static one and the $k$ integration we get
\begin{equation}\begin{split}\label{MagCP}
\Delta E_M&=-\frac{1}{2}\alpha_M^B\langle H^2(\rv_B)\rangle_A
=\frac{7\hbar c}{4\pi}\alpha_E^A\alpha_M^B\Bigl(\frac{1}{r_0^7}+\frac{1}{r_1^7}\Bigl) \\
&-\frac{16\hbar c\alpha_E^A\alpha_M^Br_0\cos \theta_0 r_1 \cos
\theta_1}{\pi r_0^3r_1^3(r_0+r_1)^5}\Bigl[r_0^2+5r_0r_1+r_1^2\Bigr].
\end{split} \end{equation}
We note that, similarly to the electric case, the magnetic interaction contains three terms: two are related to the atom-atom (depending from $r_0$ only) and atom-image (depending from $r_1$ only) interactions, and the other one depends on both coordinates $r_0$ and $r_1$. It should be also noted that the magnetic Casimir-Polder interaction between the two atoms is repulsive. However, in ordinary situations magnetic Casimir-Polder interactions are much weaker than electric ones, the total force being attractive. For large atom-wall distance ($r_1 \gg r_0$), Eq. \eqref{MagCP} reduces to the well-know stationary electric-magnetic Casimir-Polder interaction between two atoms in free space \cite{FS68,FS70}. Finally, we wish to stress that the atom-atom dynamical Casimir-Polder interactions discussed in this Section add, of course, to the known dynamical atom-wall interaction energies, discussed in \cite{SHP03,VP08,HRS04} for the case of electric interactions.

\section{Conclusions}

In this paper we have calculated the time-dependent electric and magnetic Maxwell field operators in the Heisenberg representation for a two-level atomic source near a boundary condition such as a perfectly conducting wall. The method used consists of an iterative solution of the Heisenberg equations for the field operators. The results obtained for the field operators are quite general, not depending on the specific atom-field initial state.  We have explicitly obtained solutions up to the second order in the electric charge and showed that they can be separated into a \emph{direct} and \emph{reflected} term. The latter  is related to the presence of the wall. Both terms propagate in a strictly causal way. We have
then evaluated the time-dependent electric and magnetic energy densities around the atom for a specific initial state of the system, that is the atom in its bare ground state and the field in its vacuum state. This has finally allowed us to evaluate the dynamical (time-dependent) Casimir-Polder interaction energy between the ``source'' atom and a second atom, considered as a polarizable body, when both are placed in the vicinity of the conducting wall. Both electric and magnetic Casimir-Polder energies have been considered. We have also shown that for $t \to \infty$ both the
electric and magnetic dynamical Casimir-Polder energies settle to time-independent values, which coincide with those that can be obtained by a time-independent approach. We stress that our method is general and not limited to a specific initial state of the system; in fact our solutions for the field operators in the presence of the wall can be used for any atom-field initial state, being operator solutions.

\appendix*
\section{Polarization sum}

In this Appendix we evaluate the following expressions which have been used in Section \ref{Sec:3},

\begin{equation}\begin{split}\label{Polariz}
&\sum_{j}\int d\Omega(\fkj(\rv))_{p}(\fkj(\rv_A))_{q}\\
&\sum_{j}\int d\Omega(\nabla\times\fkj(\rv))_{p}(\fkj(\rv_A))_{q}
\end{split}\end{equation}
where $p,q=x,y,z$. In order to show the method used, we first make a specific choice of $p$ and $q$, namely $p=x$ and $q=y$, and we
calculate the first expression of \eqref{Polariz}. The technique is the same for all the other cases. Using the mode functions \eqref{Funzmodoparete} and applying the sum rule for the polarization unit vectors
\begin{equation}\sum_j(\ekj)_p(\ekj)_q=\delta_{pq}-\hat{k}_p\hat{k}_q\end{equation}
where $\delta_{pq}$ is the Kronecker delta function, we get
\begin{equation}\begin{split}
&\sum_{j}\int
d\Omega(\fkj(\rv))_{x}(\fkj(\rv_A))_{y}\\
&\quad=-\frac{8}{k^2}\int d\Omega\,k_xk_y\cos\Bigl[k_{x}\Bigl(x+\frac L2\Bigr)\Bigr]\sin(k_{x}L/2)\\
&\quad\times\sin\Bigl[k_{y}\Bigl(y+\frac L2\Bigr)\Bigr]\cos(k_{y}L/2)\sin(k_zz)\sin(k_zd).
\end{split}\end{equation}
We can now exploit the fact that in the limit $L\to+\infty$, the oscillating functions containing $L$ can be replaced by their mean value over a period; for example
\begin{equation}\begin{split}
\sin^2(k_xL/2)&\to1/2\\
\sin(k_xL/2)\cos(k_xL/2)&\to0.
\end{split}\end{equation}
Thus we get
\begin{equation}\begin{split}
&\frac{2k_xk_y}{k^2}\sin(k_xx)\sin(k_yy)\sin(k_zz)\sin(k_zd)\\
&=\frac{2}{k^2}\nabla_x\nabla_y\Bigl[\cos(k_xx)\cos(k_yy)\sin(k_zz)\cos(k_zd)\Bigr].
\end{split}\end{equation}
Finally, using
\begin{equation}
\int d\Omega\cos(k_x a)\cos(k_y b)\cos(k_z c)=4\pi\frac{\sin(kD)}{kD}
\end{equation}
with $D=\sqrt{a^2+b^2+c^2}$, we get
\begin{equation}\label{Finapola}
\sum_{j}\int
d\Omega(\fkj(\rv))_{x}(\fkj(\rv_A))_{y}=\frac{4\pi}{k^3}\sum_{\nu=0,1}\nabla_x\nabla_y\frac{\sin(k
r_{\nu})}{r_{\nu}}
\end{equation}
where $r_0=\sqrt{x^2+y^2+(z-d)^2}$ and $r_1=\sqrt{x^2+y^2+(z+d)^2}$. All the other terms can be obtained using the same method. The general results are
\begin{equation}\begin{split}
&\sum_{j}\int
d\Omega(\fkj(\rv))_{p}(\fkj(\rv_A))_{q}\\
&\quad=\frac{4\pi}{k^3}\sum_{\nu = 0}^{1} (-1)^{\nu(1-\delta_{qz})}
F_{pq}\frac{\sin(k r_{\nu})}{r_{\nu}}
\end{split}\end{equation}
and
\begin{equation}\begin{split}
&\sum_{j}\int
d\Omega(\nabla\times\fkj(\rv))_{p}(\fkj(\rv_A))_{q}\\
&\quad=\frac{4\pi}{k}\sum_{\nu = 0}^{1}
(-1)^{\nu(1-\delta_{qz})}G_{pq}\frac{\sin(k r_{\nu})}{r_{\nu}}
\end{split}\end{equation}
where $F_{pq}$ and $G_{pq}$ are the differential operators defined in Section \ref{Sec:3}.

\begin{acknowledgments}
R.V. acknowledges Finnish CIMO and Turun Yliopistos\"a\"ati\"o for
financial support. R.M. and R.P. acknowledge partial financial
support from Ministero dell'Universit\`{a} e della Ricerca
Scientifica e Tecnologica and by Comitato Regionale di Ricerche
Nucleari e di Struttura della Materia.
\end{acknowledgments}

\end{document}